\documentstyle[12pt,aaspp4,flushrt,psfig]{article}
\newcommand{\beq}{\begin{equation}}
\newcommand{\beqa}{\begin{eqnarray}}
\newcommand{\eeq}{\end{equation}}
\newcommand{\eeqa}{\end{eqnarray}}

\newcommand{\etal}{{\it et al. }}

\begin{document}
\title{ 
Gravitational Lens Statistics and The Density Profile of Dark Halos
}
\author{Ryuichi Takahashi and Takeshi Chiba}
\affil{Department of Physics, Kyoto University, Kyoto 606-8502, Japan}


\begin{abstract}
We investigate the influence of the inner profile of lens objects on 
 gravitational lens statistics taking into account of
 the effect of magnification bias and both the evolution and 
the scatter of halo profiles. We take the dark halos as the lens 
objects and consider 
the following three models for the density profile of dark halos; 
SIS (singular isothermal sphere), the NFW (Navarro Frenk White) profile, 
and the generalized NFW profile which has a different slope at smaller radii.  
The mass function of dark halos is assumed to be given by the Press-Schechter
 function. We find that magnification bias for the NFW profile is order
 of magnitude larger than that for SIS. 
We estimate the sensitivity of the lensing probability of distant sources
 to the inner profile of lenses and to the cosmological parameters.
It turns out that the lensing probability is strongly dependent on the inner 
 density profile as well as on the cosmological constant.
We compare the predictions with the largest observational sample,
 the Cosmic Lens All-Sky Survey.
The absence or presence of  large splitting events in larger surveys 
currently underway such as the 2dF and SDSS 
could set constraints on the inner density profile of dark halos.  
\end{abstract}


\section{Introduction}
It has been known since the 1930s that dark matter is the gravitationally 
 main component in a variety of astrophysical objects such as galaxies and 
 clusters of galaxies, but the nature of dark matter still eludes us.
(\cite{peebles93}).  
Recently, the systematic discrepancy between numerical simulations and 
observations regarding the inner density profile of dark halos has been 
reported.
For the inner density profile of dark halos $\rho(r) \propto
 r^{-\alpha}$, numerical simulations suggest the steeper profile $\alpha
 \sim 1-1.5$, while observations suggest the shallower profile
 $\alpha \sim 0-1$.
Numerical simulations of CDM halos by Navarro, Frenk \& White 
(1996,1997, hereafter NFW)  have shown that the density profile has 
the ``universal'' form $\rho \propto r^{-1} (r+r_s)^{-2}$, 
where $r_s$ is the scale length, irrespective of the cosmological parameters, 
the initial power spectrum and the formation histories. 
Following NFW, higher-resolution simulations have been performed, 
 and the results suggest the steeper inner slope $\rho(r) \propto r^{-1.5}$
 (\cite{moore99,fm00}).
On the other hand, observations of rotation curves of the spiral and
 the low surface brightness galaxies indicate the shallower halo profile
 $\alpha \sim 0-1$ (\cite{bosch00a,blok01,bs01}).
The mass distribution of the cluster CL0024+1654 is reconstructed from 
 lensed images and indicates the flat core (\cite{tyson98}).   
It is important to resolve the discrepancy to get clues to
 the nature of dark matter.

In this regard, strong gravitational lensing effects can provide an 
important method to probe the nature of dark matter.
Strong lensing probes the inner dense region of lens objects, and
 the impact parameter is estimated as
\beq
  \xi \simeq 6.2~{h_{70}}^{-1}~~\mbox{kpc}~\frac{H_0 D_L}{0.3} \frac{\theta}
{1^{\prime \prime}}  ,
\label{xi}
\eeq
where $D_L$ is the distance to the lens and $\theta$ is the image separation.
So it is sensitive to the inner profile of lenses
 and can be a useful method for probing the inner profile of dark halos.
In fact, $\xi$ in Eq.(\ref{xi}) is comparable to the scale length $r_s$ in the
 NFW profile $\rho \propto r^{-1} (r+r_s)^{-2}$ (\cite{nfw97}). 
Dark halos is gravitationally dominant in galactic halos
 and clusters of galaxies.
However, at the inner region of galaxies, baryonic components such as bulge
 and disk are also gravitationally dominant, and dissipation processes among
 the baryons are important.
On the other hand, since clusters of galaxies are formed only 
recently,  the baryonic component distributes broadly as gas (\cite{ro77}) 
and the radial distribution of gas mass is similar to the total 
mass (\cite{ee00}). Hence, in order to study strong gravitational lensing 
 without the need to include the gravitational effects of baryonic components, 
we shall mainly concentrate on clusters of galaxies as the lens objects and 
look for large-separation images (the effect of baryons on lensing was studied 
in (\cite{pm00,kw01,keeton01})). 

In this paper, we examine the effect of the inner density profile of 
lens objects on gravitational lens statistics, including the effect of 
 magnification bias and both the evolution and the scatter of halo profiles.
Statistics of gravitational lensing of QSOs provides a useful
 tool to set constraints on the cosmological constant (\cite{ffkt92,k96}).
However, it depends on the lens model such as the profile of lenses
 and its number density as well as cosmology.
Hence, in using it as a tool to limit the cosmological constant, we must be
 careful about the uncertainties concerning the lens model.
We estimate the sensitivity of the lensing probability of distant sources
 to the inner profile of lenses and to the cosmological constant.
We consider three kinds of density profile of lens objects :
 SIS ($\rho(r) \propto r^{-2}$),
 the NFW ($\rho(r) \propto r^{-1}$ for the inner profile),
 the generalized NFW ($\rho(r) \propto r^{-\alpha}$ for the inner profile).
Only smooth and spherical models are considered. 
Subclumps (cluster galaxies) in clusters do not affect the cross section 
 of lensing (\cite{flores00}). Nonsphericity can affect the relative 
frequency of four-image lenses (\cite{rt00}).
The distribution of lenses is taken to be the Press-Schechter function.
We compare the predictions with the large observational data, 
 CLASS (the Cosmic Lens All-Sky Survey), and predict for larger surveys
 such as 2dF and SDSS.
Multiple images of large separation angles are expected to be caused by 
clusters of galaxies, so when we compare the observational data, 
we will use the large angle images ($\theta \geq 6^{\prime \prime}$).
The statistics of wide-separation lenses ($\xi \geq 10 \mbox{kpc}$) has been
 used to probe the  distributions of mass inhomogeneities derived from
 the CDM scenarios, since one does not need to be concerned with
 the physics of baryonic components and bias (\cite{wcot95}).
By studying the lensing properties in this regime assuming the CDM
 scenarios, theoretical calculations based on N-body simulations
 (\cite{cen94,wcot95,wco98}) and semi-analytical method using the
 Press-Schechter function (\cite{nw88,k95,ns97,mw00}) have been
 used to place limits on cosmological parameters.
Similarly, incorporating realistic input for mass profile and number
 density of clusters, analytical calculations have been used to set
 constraint to cosmological models (\cite{tomita96}) and lens models
 (\cite{maoz97}).
We pay particular attention to the influence of the inner lens structure
 on the statistics of wide-separation lenses (for earlier discussion before 
NFW profile, see \cite{fp96}).
                                             
Recently a similar analysis has been performed by several authors
 (\cite{wyithe00,fp01,lo00,km01}).
Wyithe, Turner \& Spergel (2001) used the generalized NFW lens model and
 suggested the optical 
 depth to multiple imaging of the distant sources is very sensitive to the 
inner lens profile, but no comparison with observational data was made. 
Li \& Ostriker (2000) found that the lensing probability is very sensitive
 to the density profile of lenses, and somewhat less so to the cosmological
 parameters such as the mean mass density in the universe and the amplitude
 of primordial fluctuations.
However, they did not take into account the scatter of dark halo profiles 
 (concentration parameter) and the uncertainty in the treatment of  
magnification bias. 
Keeton \& Madau (2001) found that the number of 
predicted lenses is strongly correlated with the core mass fraction. 
In this paper, we present a systematic study of the effect of the inner 
dark halo profile and cosmological parameters on gravitational lens 
statistics. It is very important to be careful about both the effect of 
magnification bias, which depends on the lens profile and 
magnification of each images, and both the evolution and the scatter of 
halo profiles in N-body simulations
 (\cite{bullock01}).

This paper is organized as follows.
In section 2, we briefly summarize the basic formulae of gravitational
lens statistics for various lens models.
In section 3, we examine the effect of magnification bias and compare the
 theoretical prediction with observation.
In section 4, we estimate the number of lensed images expected in
 larger surveys.
Finally, in section 5, we summarize the main results of this paper. 
We use the units of $c=G=1$.

\section{Basics}

\subsection{SIS Lens}

The SIS (singular isothermal sphere) model is frequently used in the lensing
 analysis, since it is supported by observed flat rotation curves and 
 moreover the density profile is very simple and quantities related to
 gravitational lensing can be written in simple analytic 
forms (\cite{tog84}).  
 SIS is characterized by the one-dimensional velocity dispersion $v$, and
 the density profile is $\rho (r) = {v^2}/{2 \pi}{r^2}$ .
The lens equation leads to the two solutions (image positions) at
 $x_{\pm} = y \pm 1$ with magnifications $\mu_{\pm}=\left| (y/x_{\pm})
 (dy/dx_{\pm}) \right|^{-1}= 1/y \pm 1$, where 
 dimensionless quantities $x$ and $y$ are the impact parameter
 divided by the Einstein radius in the lens plane and the source
 position divided by it in the source plane (\cite{sef92}).  
 Magnification of the brighter (fainter) image is $\mu_{+} (\mu_{-})$, 
 and the total magnification is $\mu (y)= 2/y$.
The splitting angle between the two images is $\theta = 8 \pi v^2
 {D_{LS}}/{D_S}$, where $D_L$, $D_S$ and $D_{LS}$ are the angular diameter
 distances between the observer, lens and source. 
We shall adopt the so-called filled beam distance (\cite{dr73}), since 
 the ray shooting in an inhomogeneous universe created by N-body simulations 
is consistent with it (\cite{tomita98,tah99}).

When the light ray from the source passes near the lens object, 
 if $\left| y \right| \leq 1$, double images form.
We define the cross section ${\sigma} (v,z_L,z_S)$ as the area of a 
 region in the source plane which satisfies the following criteria;
 (i)double images are formed,
 and (ii)the magnification is
 larger than the minimum amplification $\mu_{*}$.
The second condition is needed for the calculation of magnification
 bias (see Eq.(\ref{st19}) below).
Then, the cross section is given by, 
\beq
   \sigma(v,z_L,z_S) = \pi r_E^2
\times 2 \int_0^1 dy y \Theta (\mu (y) - \mu _{*}) ,
\label{st13}
\eeq
where $r_E (=4 \pi v^2 D_{LS})$ is the Einstein radius in the source plane,
 and $\Theta (x)$ is the step function.

\subsection{Generalized NFW Lens}

In this section we consider the generalized NFW profile for dark halos
 (\cite{wyithe00,lo00}), 
\beq
  \rho (r) = \frac{\rho _s}{\left( \frac{r}{r_s} \right)^{\alpha}
 \left( \frac{r}{r_s}+1 \right)^{3 - \alpha}} ,
\label{gnfw1} 
\eeq
where $\rho_s$ and $r_s$ are parameters which depend on the mass of halo $M$ 
 and redshift $z$, and $\alpha (0 \leq \alpha \leq 2)$ is a constant.
The scale radius $r_s$ is about $10$ kpc on a galactic halo scale
 ($M \sim 10^{12} M_{\odot}$) and $100$ kpc on a cluster scale
 ($M \sim 10^{15} M_{\odot}$) (\cite{nfw97}).    
Since the impact parameter $\xi$ is comparable to the scale radius $r_s$
 (see Eq.(\ref{xi})),  we could set a strong constraint on inner slope
 $\alpha$ in Eq.(\ref{gnfw1}) by using strong gravitational lensing.
We consider the case of $\alpha=0.5, 1, 1.5$.

The concentration parameter $c$ and the characteristic density $\delta_c$
 are respectively defined by
\beqa
 c &=& \frac{r_{vir}}{r_s}, \\
 \delta_c &=& \frac{\rho_s}{\rho_{m}(z)}
 = \frac{\Delta_{vir}}{3} \frac{c^3}{\int_0^c dx~x^{2-\alpha}
 (1+x)^{\alpha-3}}.
\label{gnfw2} 
\eeqa
$c$ represents degree of the mass concentration at the inner region of a halo. 
The virial radius of the halo $r_{vir}$ is related to the mass $M$ and
 redshift $z$ as
\beq
  H_0 r_{vir}=5.643 \times 10^{-5} h^{1/3} \Omega_0^{-1/3} (1+z)^{-1}
\left( \frac{\Delta_{vir}}{18 \pi^2} \right)^{-1/3} 
\left( \frac{M}{10^{12} M_{\odot}} \right)^{1/3}  ,
\label{gnfw3}
\eeq 
where $\Delta_{vir}$ is the overdensity of the halo
($\Delta_{vir}=18 \pi^2$ for EdS $(\Omega_0=1)$ model, and for other
 cosmological models we use the fitting formulae given in \cite{ks96} and
 \cite{ns97}).

 For the NFW profile ($\alpha=1$), $c$ is fitted by the numerical
 simulation (\cite{nfw97}).
For the generalized NFW profile, following the recent work 
(\cite{bullock01,km01}), we define $c$ to be $c=r_{vir}/r_{-2}$, where 
$r_{-2}$ is the radius at which the logarithmic slope of the density profile
 is $-2$ ($r_{-2}$ is equivalent to $r_s$ for the NFW profile). 
Thus we obtain  $c=(2-\alpha)^{-1} r_{vir}/r_s$. 

However, there is a scatter in the concentration parameter $c$ at fixed mass
 and redshift (\cite{bullock01,jing00}).
Keeton \& Madau (2001) pointed out the importance of the scatter of $c$
 on the lensing statistics. 
We adopt a log-normal function as the probability distribution function
 of $c$,
\beq
  p(c)dc=\frac{1}{\sqrt{2 \pi}\sigma_c} \exp \left[ -\frac{\left( \ln c
 - \ln c_{med} \right)^2}{2 \sigma_c^2} \right] d\ln c,
\label{pdf} 
\eeq
 with $\sigma_c=0.18$ (\cite{bullock01,jing00}). 

Using the toy model of a $c_{med}-M$ relation found by Bullock \etal (2001), 
we estimate the concentration of the halos,
\beq
  c_{med} = \frac{10}{1+z} \left( \frac{M}{M_*} \right)^{-\beta},
\label{cmed}
\eeq
where the fitting parameters $(M_*, \beta)$ are $(7.0 \times 10^{13}
 M_{\odot}, 0.16)$ for EdS ($h=0.5, \Omega_0=1, \lambda_0=0, \sigma_8=0.67$)
 model, $(2.1 \times 10^{13} M_{\odot}, 0.14)$ for $\Lambda$ ($h=0.7,
 \Omega_0=0.3, \lambda_0=0.7, \sigma_8=1$) model and $(2.6 \times 10^{13}
 M_{\odot}, 0.19)$ for open ($h=0.7, \Omega_0=0.3, \lambda_0=0,
 \sigma_8=0.85$) model.  
Recent high resolution numerical simulations (\cite{bullock01}) 
show that $c$ does depend on $z$ contrary to the earlier suspicious 
that $c$ does not vary much with the redshift (\cite{nfw97}).

The lens equation for the halo with the generalized NFW profile is 
\beq
   y = x-A~\frac{g( x )}{x}
\label{gnfw4}
\eeq
where $x=\xi/r_s$ ($\xi$ is the impact parameter in the lens plane),
 $y =(D_L/D_S) (\eta / r_s)$ ($\eta$ is source position in the source plane),
 and 
\beqa
   g ( x ) &=& \int_0^x dz~z^{2 - \alpha}  \int_0^{\pi / 2}
 d\theta~\frac{\cos {\theta}}{(\cos {\theta} + z)^{3 - \alpha}} ,
\label{gnfw5}  \\
   A &=& 16 \pi~\rho _s~\frac{D_L D_{L S}}{D_S}~r_s .
\label{gnfw6}
\eeqa
For $\alpha=0, 1, 2$, the integration of the Eq.(\ref{gnfw5}) can be
 carried out analytically (the lens equation for the NFW lens was obtained by
 Bartelmann (1996)).
In Fig.2, we show the lens equation for the NFW lens with $A=10$.
If $|y| \leq y_{crit}$, three images $x_i~(i=1, 2, 3, x_3 \geq x_2 
 \geq x_1 )$ form.
The image positions $x_i=x_i(y)$ with magnifications $\mu_{i}(y)$
 can be obtained numerically.
The image separation angle $\theta$ is defined as the separation between 
the outer two images and  depends on the source position $y$. We use the 
averaged value in the source plane.
\beq
   \theta (M,z_L) = \frac{1}{\pi {y_{crit}^2 }} \int_0^{y_{crit}}
 d(\pi y^2) ( x_3 - x_1 ) \frac{r_s(M,z_L)}{D_L} .
\label{gnfw8}
\eeq
The total magnification is the sum of the magnifications of
 three images ($\sum_{i=1}^3 \mu_i$) and the magnification of the fainter 
image is smaller magnification of outer two images 
(min \{ $\mu_1, \mu_3$ \}).  
Following the case of SIS (see Eq.(\ref{st13})), the cross section is
 defined by
\beq
   \sigma(M,z_L,z_S) =\pi~\left( \frac{D_S}{D_L} r_s \right)^2 
\times 2 \int_0^{y_{crit}} dy y~\Theta (\mu (y) - \mu _{*}) . 
\label{gnfw9}   
\eeq

\subsection{Lensing Probability}

The lensing probability for a source at redshift $z_S$ is (\cite{sef92})
\beq
   P (z_S) = \int_0^{\infty} dM \int_0^{z_S} dz_L~\frac{1}{{D_S}^2}
~\frac{(1+z_L)^2}{H(z_L)} {D_L}^2~\sigma (M,z_L,z_S)~N_M (M,z_L),
\label{gnfw10}    
\eeq
where $H(z) = H_0 \left[~\Omega_0 (1+z)^3 +(1-\Omega_0-\lambda_0)
 (1+z)^2 +\lambda_0~\right]^{1/2}$ is the Hubble parameter at redshift $z$,
 and $N_M (M,z_L)$ is the comoving number density of lenses and
 is assumed to be given by the Press-Schechter mass function.
The one-dimensional velocity dispersion $v$ is related to the mass through
 $v = \left( {M}/{2 r_{vir}} \right) ^{1/2}$, and this relates the PS mass
 function to the SIS lens profile.
The probability of the image separation angle is obtained  by computing
 $dP/d\theta$.

Since gravitational lensing causes a magnification of images, lensed sources
 are over-represented in a magnitude-limited sample and 
 the actual lensing probability is enhanced.
This selection effect is called magnification bias.
Let $\Phi _S (z_S,L)~dL$ be the luminosity function of sources.
The observed flux $S$ for a lensed source is related to the luminosity $L$,
   $L=4 \pi (1+z_S)^4~D^2_S~(1+z_S)^{\gamma - 1} S$,
where the factor $(1+z_S)^{\gamma - 1}$ is the K-correction, which assumes
 that the energy spectrum of source is of the form $E \propto \nu^{-\gamma}$.
When one searches for lensed source of the observed flux $S$, the lensing 
 probability increases as
\beq
   P^{B} (z_S,L) = \frac{1}{\Phi _S (z_S,L)} \int_1^{\infty} d\mu _{*}
~\left| \frac{d}{d\mu_{*}} P (z_S) \right|~\Phi _S (z_S,L / \mu _{*})
~1/ \mu_{*} .
\label{st19}
\eeq

Including magnification bias, the lensing probability $P^B (z_S,L)$
 for the generalized NFW lens
 is expressed as Eq.(\ref{gnfw10}) with
 $\sigma(M,z_L,z_S)$ replaced by
\beq
  \sigma^B(M,z_L,z_S,L) = \pi~\left( \frac{D_S}{D_L} r_s \right)^2
 \times~\frac{2}{\Phi _S (z_S,L)} \int_0^{y_{crit}} dy~y
~\Phi _S (z_S,L / {\mu (y)}) / {\mu (y)}.  
\label{bias}
\eeq  
$\mu (y)$ can be the magnification of the total images or the
 fainter image. 
We will discuss each case in Sec. 3.1.

\subsection{Press-Schechter Function}
We use the Press-Schechter (PS) function (\cite{ps74}) for computing
 the number density of the halos. 
The PS mass function is given by
\beq
  N_M (M,z)~dM = \sqrt{\frac{2} {\pi}}~\frac{\rho_0}{M}
~\frac{\delta _{crit} (z)}{\sigma (R)}~\left| \frac{d\ln \sigma}{d\ln M} 
\right|~\exp \left[ - \frac{{\delta}^2 _{crit} (z)}
{2 {\sigma}^2 (R)} \right]~\frac{dM}{M} .
\label{ps1}
\eeq
Here $\rho_0$ is the mean mass density of the universe at present and
 $\sigma(R)$ is the linear density fluctuation presently on the
 comoving scale $R$,
\beq
  {\sigma}^2 (R) = \frac{1}{2 {\pi}^2} \int_0^{\infty} dk~k^2~P(k)
~W^2 (kR) 
~,~~~R = \left( \frac{2 M}{\Omega _0 {H_0}^2} \right) ^{1/3} ,
\label{ps2}
\eeq
where $P(k)$ and $W(kR)$ are the power spectrum at present
 (\cite{bbks86,sugiyama95}) and the top-hat window function. 
We normalize $\sigma(R)$ so that $\sigma(R=8h^{-1}{\rm Mpc})=\sigma_8$, and 
the critical density contrast is $\delta_{crit}(z)=1.686/{D_1(z)}$,
 where $D_1(z)$ is the linear growth rate normalized to unity at $z=0$.

\section{Results}
In order to explore the dependence of the lensing probability on cosmological
 models, we consider three representative models;
 EdS model ($h=0.5, \Omega_0=1, \lambda_0=0, \sigma_8=0.67$),
 $\Lambda$ model ($h=0.7, \Omega_0=0.3, \lambda_0=0.7, \sigma_8=1$),
 open model ($h=0.7, \Omega_0=0.3, \lambda_0=0, \sigma_8=0.85$).

\subsection{The Effect of Magnification Bias}

We demonstrate the effect of  magnification bias by calculating the biased 
 lensing probability Eq.(\ref{gnfw10}, \ref{st19}). 
In calculating magnification bias, we use an optical QSO luminosity 
function. We adopt the 2dF QSO redshift survey data which include about 6000
 QSOs with the redshift distribution $0.35 < z < 2.3$ (\cite{boyle00}).  
The QSO luminosity function is fitted by the two-power-law model 
\beqa
   \Phi _S (z_S,L) dL &=& \frac{\Phi_S^{*}~dL}{(L/L_{*})^{c_1}
 + (L/L_{*})^{c_2}}, \label{st19-1} \\ 
   L_{*} (z_S) &=& L_{0 *} 10^{k_1z_S+k_2z_S^2} .
\label{st20}
\eeqa
The fitting parameters for $\Lambda$ model are given by 
\beqa
 (c_1, c_2, M_{B *}, k_1, k_2)
 &=& (3.41, 1.58, -21.14+5 \log h, 1.36, -0.27), \\
 \Phi_S^{*} &=& 2.88 \times 10^{-6} h^3
 \mbox{Mpc}^{-3} \mbox{mag}^{-1},
\eeqa
where $M_{B *}$ is the absolute B-band magnitude corresponding 
to $L_{0 *}$ (Eq.(\ref{st20})).
We take the absolute magnitude of a source is $M_B=-25.8$ mag in
 Eq.(\ref{st20}) and consider the image separation range $\theta
 \geq 0.3^{\prime \prime}$.
For the energy spectrum index of source (see section 2.3),
 we use $\gamma=0.5$ for the optical QSOs (\cite{boyle88}). 
In Fig.2, the effect of magnification bias for SIS and the NFW is shown 
 for $\Lambda$ model. 
The vertical axis is the lensing probability with magnification bias divided
 by that without it. 
As the source redshift is higher, the amplitude of magnification bias
 is smaller. This is because the number of fainter QSOs 
(its luminosity is $L \leq L_{*}$) is smaller at higher redshift 
(Eq.(\ref{st19-1})), so the integration of luminosity function 
Eq.(\ref{bias}) takes a smaller value.   
{}From Fig.2, we find that the magnification bias effect for the NFW
 is order of magnitude larger than that for SIS.
This is due to the fact that the magnification for the NFW is divergent at 
 $y=y_{crit}$ and $y=0$ (see Fig.2).
 Hence, if we attempt to predict lensing 
frequencies by using the NFW lens model, magnification bias should not 
be ignored. 

In Fig.2, we also compare the case when $\mu$ in Eq.(\ref{bias})
 is the magnification of the total images with the case
 of the fainter image.
Depending on the properties the gravitational lensing configuration, 
 we should use $\mu$ in the bias factor as the magnification of the total
 images (we call ``$\mu$ total'') or the fainter image among the outer 
two images (``$\mu$ fainter''). If individual sources in a sample are not 
examined closely enough to
 determine whether they are lensed or not, the magnification of the 
 fainter image should be used (\cite{st93,cen94}) because the fainter image
 should be bright enough to be recognized as one of the multiple images.
On the contrary, if one searches for lensed source of small separation
 angles, then the total magnification may be relevant, because it is likely
 that the brightness of a lensed source with a small separation is
 recognized as the total brightness of all the images. 
So as far as images of large separation angles are concerned, 
 it may be better to use $\mu$ as the magnification of the fainter image.   

{}From Fig.2, different choice of $\mu$ in the bias
 factor greatly changes the amplitude of magnification bias for SIS, 
 but does not change it so much for the NFW; the difference is only factor 
of three. This is because the magnification for the NFW is divergent at 
 $y=0$ more strongly than at $y=y_{crit}$ and the magnifications of 
both of the outer two images (including the fainter image) are divergent 
at $y=0$ (see Fig.1).
In the following we will consider both cases ($\mu$ total or $\mu$ fainter), 
since the degree of the magnification bias depends on the method of 
gravitational lensing search and thus the expected number of lensed sources 
will be in between.

\subsection{Comparison with the CLASS Data}
In this section, we compare predicted lens statistics with a well-defined
 observational sample.
The Cosmic Lens All-Sky Survey (CLASS) is the largest statistically
 homogeneous search for gravitational lenses (\cite{bm00}).
The sample comprises 10,499 flat-spectrum radio sources with flux $S >
 30$ mJy at 5 GHz, and includes 18 gravitational lenses with image
 separations $0.3^{\prime \prime} \leq \theta \leq 6.0^{\prime \prime}$.
An explicit search for lenses with image separations $6.0^{\prime \prime} \leq
 \theta \leq 15.0^{\prime \prime}$ has found no lenses (\cite{phillips00}).
The flux distribution of flat-spectrum radio sources in the CLASS samples 
can be described as a power-law (\cite{rt00})
\beq
\Phi_S(z_S,L) dL \propto L^{-2.1} dL.
\eeq
It is a steeper number-flux relation than predicted by the Dunlop \& Peacock
 (1990) luminosity function (whose slope is $-1.8$ for the faint sources).
The redshift distribution of the full CLASS sample is not known.
Marlow \etal (2000) reported the redshifts for a small subsample of 27 sources.
We assume that the redshift distribution of the full sample
 is identical to that of the subsample.
We need not consider the magnitude distribution of the sample, since 
 the lensing probability does not depend on the magnitude of the
 source for the power-law luminosity function.

In Fig.3-5, we show the predicted image separation distribution of expected
 number of lensed source's in the CLASS with magnification bias for each case
 of $\mu$.
The angular resolution of parent survey in the CLASS is very low. 
The survey flux encompasses all the flux of even the widest separation 
lenses observed to date. The CLASS survey then reimages the systems 
looking for multiple imaging (\cite{myers95,mw00,helbig00}). 
Hence, for the lensed source, $\mu$ is given by summing the fluxes of all
 images (i.e. ``$\mu$ total'' is appropriate). 
In these figures, we also show the case of ``$\mu$ fainter '' for comparison. 
We note that the lenses of large separation angle ($\theta \geq
 6^{\prime \prime}$) is expected to be lensed by clusters of galaxies
 (or equally dark halos), so we should compare the theoretical prediction
 with large image separation side in CLASS data.
In Fig.3, we take the NFW profile and compare the dependence
 on the cosmological models.
The lensing probability is the highest for EdS model.
This is due to the fact that the PS mass function is proportional to
 $\Omega_0$ (see Eq.(\ref{ps1})) and the concentration parameter $c$ is
 the highest (see Eq.(\ref{cmed})). 
In Fig.4, we show the lens model dependence for $\Lambda$ model.
Since there are no lenses for large image separations ($6^{\prime \prime}
 \leq \theta \leq 15^{\prime \prime}$) in CLASS sample,  
the steeper inner profile ($\alpha > 1.5$) seems disfavored. 
However, it is preliminary, since we do not know whether the subsample 
by Marlow et al. is a fair sample. 

We estimate the sensitivity of the lensing probability to the model
 parameters. 
Using Eq.(\ref{gnfw10},\ref{st19}) and the CLASS data with
the magnification of total images, it is estimated around $\alpha=1,
 \sigma_8=1, \lambda_0=1-\Omega_0=0.7, w=-1$
\beqa
 \left. \frac{\delta N_{\theta}}{N_{\theta}} \right|_{\theta=6^{\prime
 \prime}} \simeq 
 7.4~\frac{\delta \alpha}{\alpha}  
+ 5.7~\frac{\delta c_{med}}{c_{med}}
+ 4.3~\frac{\delta \sigma_8}{\sigma_8} 
-5.7~\frac{\delta \lambda_0}{\lambda_0} 
+ 0.35~\frac{\delta w}{w} ,
\label{lnfw21} \\
 \left. \frac{\delta N_{\theta}}{N_{\theta}} \right|_{\theta=12^{\prime
 \prime}} \simeq 
 7.3~\frac{\delta \alpha}{\alpha}
+ 6.3~\frac{\delta c_{med}}{c_{med}}
+ 6.1~\frac{\delta \sigma_8}{\sigma_8} 
-6.3~\frac{\delta \lambda_0}{\lambda_0} 
+ 0.27~\frac{\delta w}{w}   ,
\label{lnfw22} 
\eeqa
where $w$ is the equation of state of dark energy ($w=-1$ for the 
cosmological constant) and a flat FRW model is assumed. 
Eqs.(\ref{lnfw21}) and (\ref{lnfw22}) indicate clearly that the lensing 
probability is very sensitive to the lens model parameters ($\alpha,c_{med}$)
 as well as the cosmological parameters ($\lambda_0,\sigma_8$),
but not sensitive to dark energy parameter ($w$). 
For example, in order to put constraint on  $\lambda_0$ within 
${\cal O}(10)\%$ accuracy, one needs to determine the inner profile $\alpha$ 
and the concentration parameter $c$ with similar accuracy. 
The dispersion of concentration parameter $c$ is about $0.2$
 (\cite{jing00,bullock01}). However, the current uncertainty in $\alpha$ is 
${\cal O}(50)\%$ ($\alpha \sim 1 - 1.5$).
Hence, it may be more useful to use the number of  
large image separation lenses to constrain the inner density profile $\alpha$. 
Cosmological parameters $\lambda_0$ and $\sigma_8$ could be determined 
 within ${\cal O}(10)\%$ by using CMB, SNIa and number count of clusters 
data (\cite{bernardis01,fc01}). The sensitivity of the lensing probability 
to the model parameters was also estimated by Li \& Ostriker (2000), 
but they assumed a single source at $z=1.5$ and did not include the effect of
 magnification bias. 
So the detailed comparison  may not be so meaningful. However, we note that 
the dependence on $c$ and the cosmological constant parameter 
is slightly larger than that found by Li \& Ostriker (2000).

In Fig.5, we show the effect of the scatter in N-body simulation on the 
 image separation distribution.
The dispersion of $N_{\theta}$, $\sigma_{\theta}^2$, is calculated by using 
 the probability distribution function of $c$ (Eq.(\ref{pdf})).
We use the NFW profile for $\Lambda$ model and compare the amplitude of the
 square of the dispersion $\sigma_{\theta}^2$ with the predicted number of 
lenses $N_{\theta}$. We find that the amplitude of $\sigma_{\theta}$ 
is comparable to or larger than $N_{\theta}$, so the scatter of halo 
profiles strongly affect the lensing probability. When compared with 
Fig.3, we also find that the scatter is too large to distinguish the
 different cosmological models, even if the lens model ($\alpha$) is fixed.

\section{Prediction for Future Survey}
With the current data, the constraints on the parameters are not sufficient.
However, we expect that larger surveys currently underway such as the 2dF and 
 SDSS detect a larger number of  lenses.
For example, the Sloan Digital Sky Survey (SDSS) plans a spectroscopic survey
 of $10^5$ QSOs over $\pi$ steradian brighter than $i^{\prime} \sim 19$ at
 $z \leq 3.0$; at redshift between 3.0 and about 5.2, the limiting magnitude
 will be $i^{\prime} \sim 20$ (\cite{york00}).
In this section, we will make predictions for the SDSS.   

Let $N_{\theta}d\theta$ be the expected number of lensed QSOs with image
separation $\theta \sim \theta+d\theta$ within solid angle $\pi$ in the sky. 
We use the QSOs luminosity function $\Phi_S$ for $z \geq 3$ in SDSS
 data (\cite{fan01}), since $\Phi_S$ from 2dF redshift survey is known only
for lower redshift QSOs ($z \leq 3$) (\cite{boyle00}). 
 $\Phi_S$ is fitted by a power-law, 
\beq
  \Phi_S (z_S,M_{1450}) = \Phi_S^{*} 10^{-0.4 \left\{ M_{1450}+26-\alpha
 \left( z_S-3 \right) \left( \beta+1 \right) \right\} },
\eeq
where $M_{1450}$ is the absolute AB magnitude of the quasar continuum 
at $1450 \AA$ in the rest frame.
We assume $m_{i^{\prime}}=m_{1450}+0.7$.
The fitting parameters for $\Lambda$ model are given by (\cite{fan01}) 
\beq
  (\Phi_S^{*},\alpha,\beta) = (2.6 \times 10^{-7} h^3 \mbox{Mpc}^{-3} 
 \mbox{mag}^{-1}, 0.75, -2.58).
\eeq
Similarly, for lower redshift sources ($z \leq 3$) the QSO luminosity
 function (Eq.(\ref{st19-1},\ref{st20})) is used (\cite{boyle00}).  
Then, using the QSO luminosity function $\Phi_S$, $N_{\theta}(\theta)$ can be
 calculated as
\beqa
  N_{\theta}(\theta) &=& \int_{0}^{z_{max}} dz \frac{dV_{\pi}}{dz}
 \int_{L_{lim}(z)}^{\infty} dL~\frac{dP}{d\theta} (\theta,z,L)
 \Phi_S (z,L) ,\\
 \frac{dV_{\pi}}{dz} &=& \pi \frac{ (1+z)^2 D(z)^2}{H(z)}, \nonumber 
\eeqa
where $z_{max}=5.2$ and $L_{lim}(z)$ is calculated from the limiting magnitude.
In Fig.6, we show the predicted image separation distribution for SDSS.
We use various lens model for $\Lambda$ model. 
In this model, the number of QSOs is  expected to be about 26,000. 
In Table 1, the expected number of large image separation lenses
 ($6^{\prime \prime} \leq \theta \leq 30^{\prime \prime}$) is shown. 
We find that the ambiguity resulting from the treatment of magnification 
bias is not so large. {}From this table, we expect that the future SDSS 
data could set constraint on the inner density profile.

\section{Summary and Discussion} 
We have examined the influence of the inner density profile of lenses on
 gravitational lens statistics carefully taking into account of the effect
 of magnification bias and the evolution and the scatter in halo profiles.
We have estimated the sensitivity of the lensing probability 
 to the inner density profile of lenses and to the cosmological constant.
We have found that lensing probability is strongly dependent on the inner
 density profile as well as on the cosmological constant.
We have also shown that magnification bias for the NFW is order of magnitude 
 larger than that for SIS. There is an uncertainty in the treatment of 
magnification bias: fainter image should be detected in the survey, or  
 the light from both the fainter and brighter images is initially unresolved 
in a single image and thus the total image should be detected.
However, for NFW profile, difference between 
the magnification bias of the fainter image and that of the total image is 
found to be only by factor of three. 
In any case, we should be careful about magnification bias which strongly 
depends on the  lens profile.
We have compared the predictions with the  CLASS data and suggested that 
the steeper inner profile ($\alpha > 1.5$) seems disfavored. 
The absence or presence of  large splitting events in larger surveys 
currently underway such as the 2dF and SDSS 
could set constraints on the inner density profile of dark halos.  

Recently, using the arc statistics of gravitational lensing, various authors
 have examined the inner profile of dark halos
 (\cite{bartelmann98,mh00,oguri01}). 
Comparing  with the existing observational data,
Molikawa \& Hattori (2001) and Oguri, Taruya \& Suto (2001) suggested that
 the steeper inner profile of
 dark halos ($\alpha >1$ or even $\alpha > 1.5$) is favored. 
On the other hand,  the absence of large images separations in 
QSOs multiple images in the CLASS sample constrains the inner profile and
rather disfavors the steeper profile. 
Combining the arc statistics and the statistics of QSOs multiple image, 
we could narrow the allowed range of the inner profile of 
dark halos, or more interestingly both methods might exhibit discrepancy. 

In any event, larger surveys will produce a lot of QSOs multiple 
images in near future and theoretical development especially concerning 
the uncertainties of various models will be expected.
So we will get clues to the nature of dark matter.

\acknowledgments
We would like to thank Professor Yasushi Suto, Dr. Ryoichi Nishi 
for useful discussion and comments. One of the authors (RT) also thanks 
Dr. Atsushi Taruya and Dr. Masamune Oguri for useful comments. 
This work was supported in part by a Grant-in-Aid for Scientific 
Research (No.13740154) from the Japan Society for 
the Promotion of Science.

\clearpage

\begin{figure}
  \begin{center}
  \epsfxsize=120mm
  \epsfbox{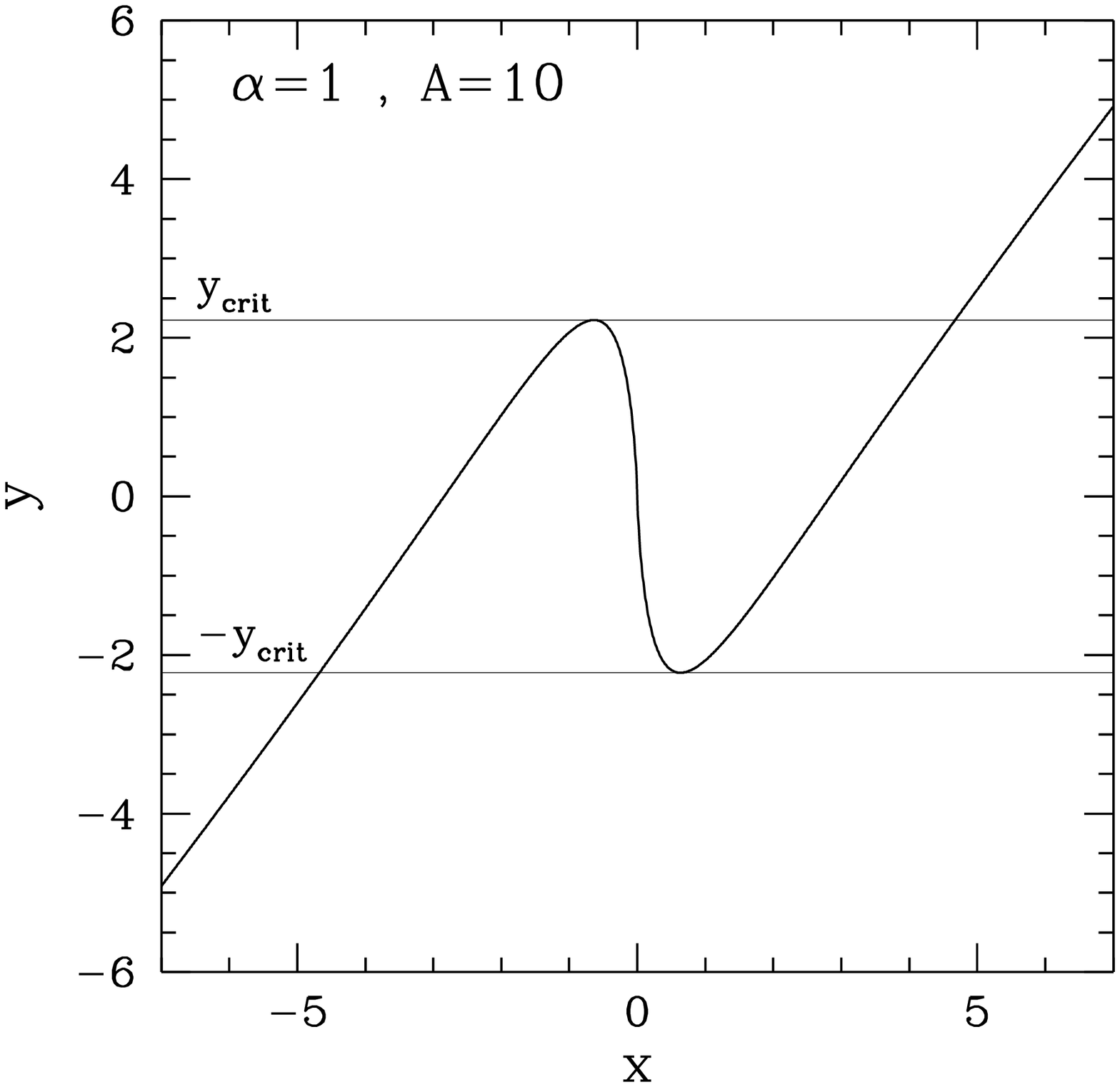}
  \end{center}
\caption{The lens equation for the NFW.
 The case with $A=10$ is shown.
 The horizontal axis is $x$, which is the impact parameter normalized 
 by a scale radius $r_s$ in the lens plane; 
 the vertical axis is $y$, which is the source position normalized by a scale 
 radius $r_s$ in the source plane.
 Multiple images are formed when $\left| y \right| \leq y_{crit}$.}

\end{figure}
\begin{figure}
  \begin{center}
  \epsfxsize=120mm
  \epsfbox{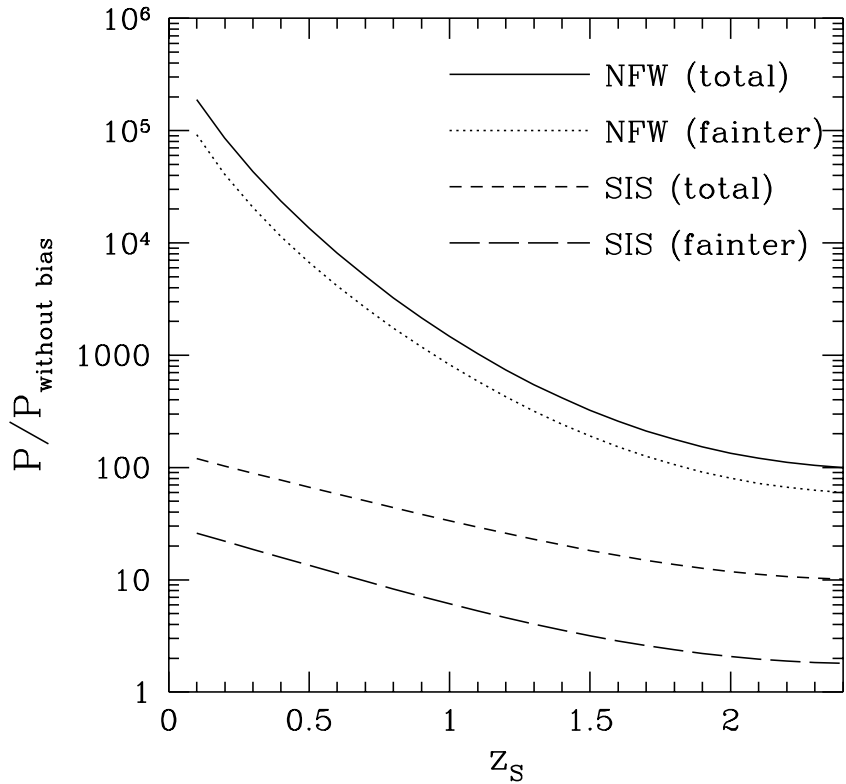}
  \end{center}
\caption{The amplitude of magnification bias for $\Lambda$ model
 $(h=0.7, \Omega_0=0.3, \lambda_0=0.7, \sigma_8=1.0)$ with the absolute
 magnitude of a source being $M_B=-25.8$ mag.
 The horizontal axis is $z_S$, which is the source redshift; the vertical
 axis is a ratio lensing probability with the magnification bias to
 that without it. Image separation range  $\theta \geq
 0.3^{\prime \prime}$.
 The solid (dotted) line is the NFW profile for the case of the 
magnification of the total images (the fainter image). 
 The short (long dashed) line is SIS for the case of the magnification 
of the total images (the fainter image). }

\end{figure}

\begin{figure}
  \begin{center}
  \epsfxsize=120mm
  \epsfbox{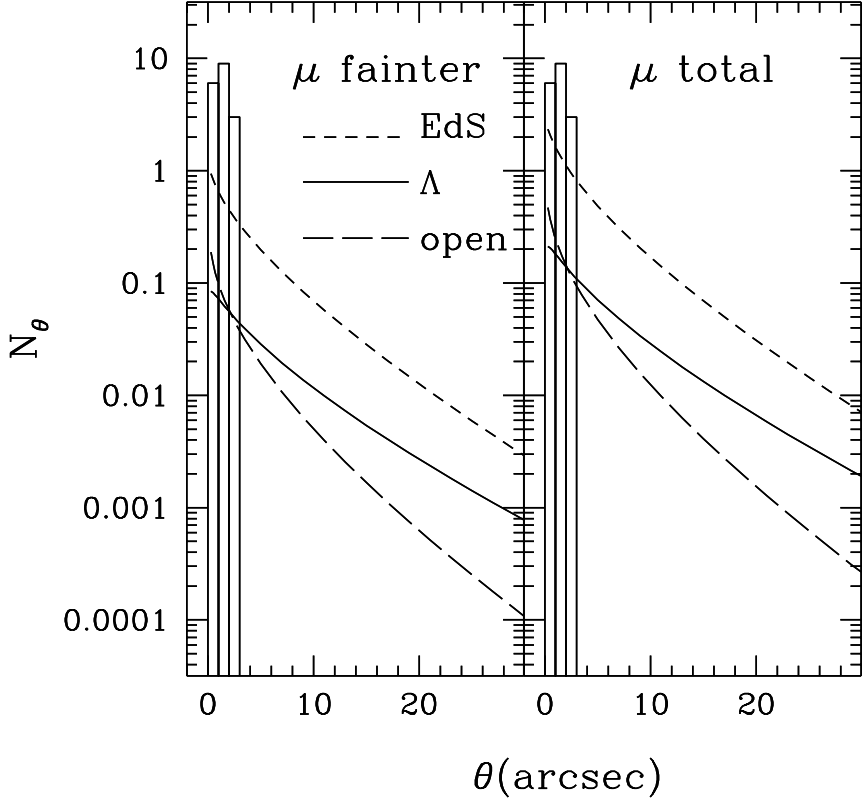}
  \end{center}
\caption{The distribution of image separations for the NFW lens model. 
The assumed cosmologies are 
EdS model(short-dashed: $h=0.5, \Omega_0=1, \lambda_0=0, \sigma_8=0.67$), 
 $\Lambda$ model (solid: $h=0.7, \Omega_0=0.3, \lambda_0=0.7, \sigma_8=1$) and 
 open model(long-dashed: $h=0.7, \Omega_0=0.3, \lambda_0=0, \sigma_8=0.85$).
The left figure is for the selection condition in which 
fainter image should be detected in the survey, and 
the right figure is for the selection condition that the light from 
both the fainter and brighter images is initially unresolved in a single 
image. ``$\mu$ total'' is appropriate in the CLASS. 
The observational data from CLASS are shown by the histogram.}

\end{figure}
\begin{figure}
  \begin{center}
  \epsfxsize=120mm
  \epsfbox{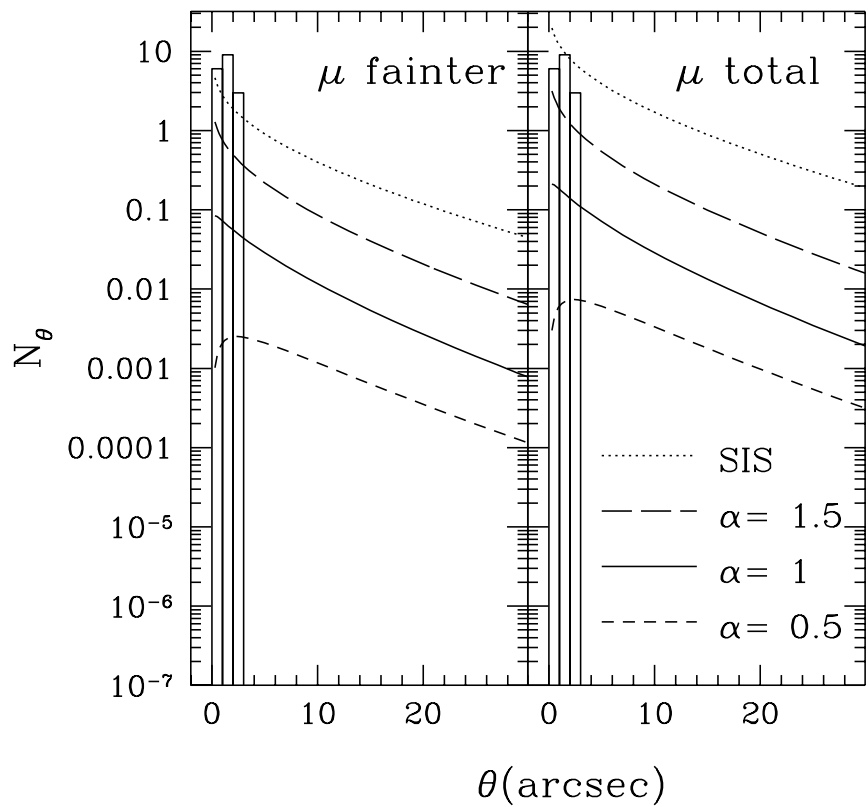}
  \end{center}
\caption{Same as Fig.3, but for various lens models (dotted: SIS, 
long-dashed: the generalized NFW with $\alpha=1.5$,  
solid: NFW ($\alpha=1$), short-dashed: the generalized NFW 
with $\alpha=0.5$).  $\Lambda$ model is assumed for cosmology. 
 The observational data from CLASS are shown by the histogram.}

\end{figure}
\begin{figure}
  \begin{center}
  \epsfxsize=120mm
  \epsfbox{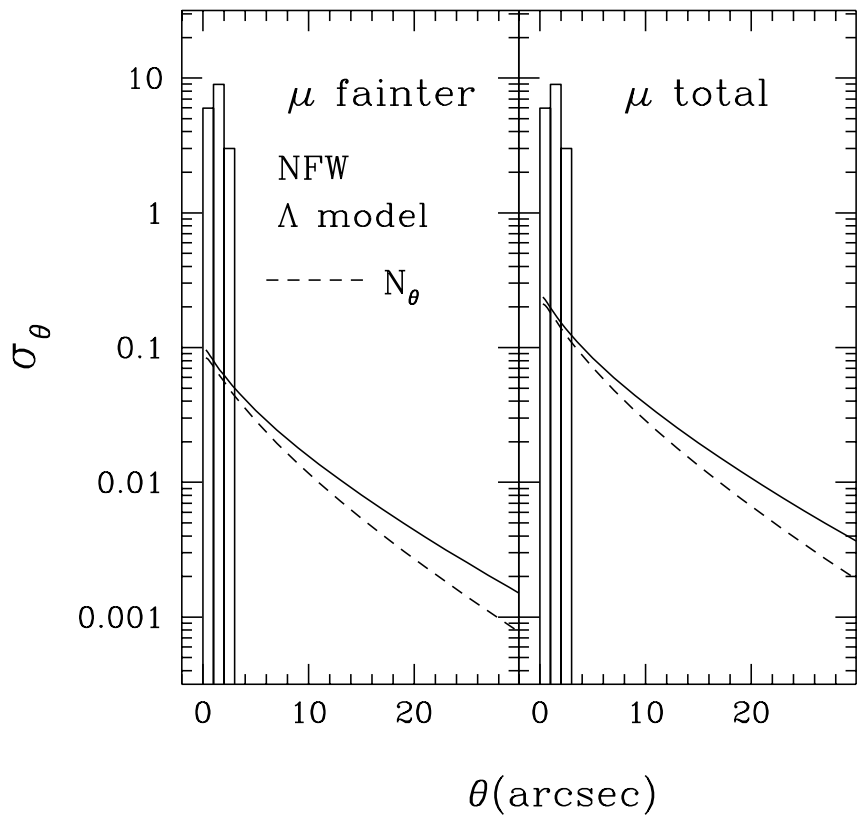}
  \end{center}
\caption{The dispersion of the predicted number of lenses caused by 
the scatter of halo profiles in N-body simulation.
The solid line is the square of the dispersion (standard deviation), 
the dashed line is the averaged distribution of image separations.
$\Lambda$ model and the NFW profile ($\alpha=1$) are assumed.}

\end{figure}
\begin{figure}
  \begin{center}
  \epsfxsize=120mm
  \epsfbox{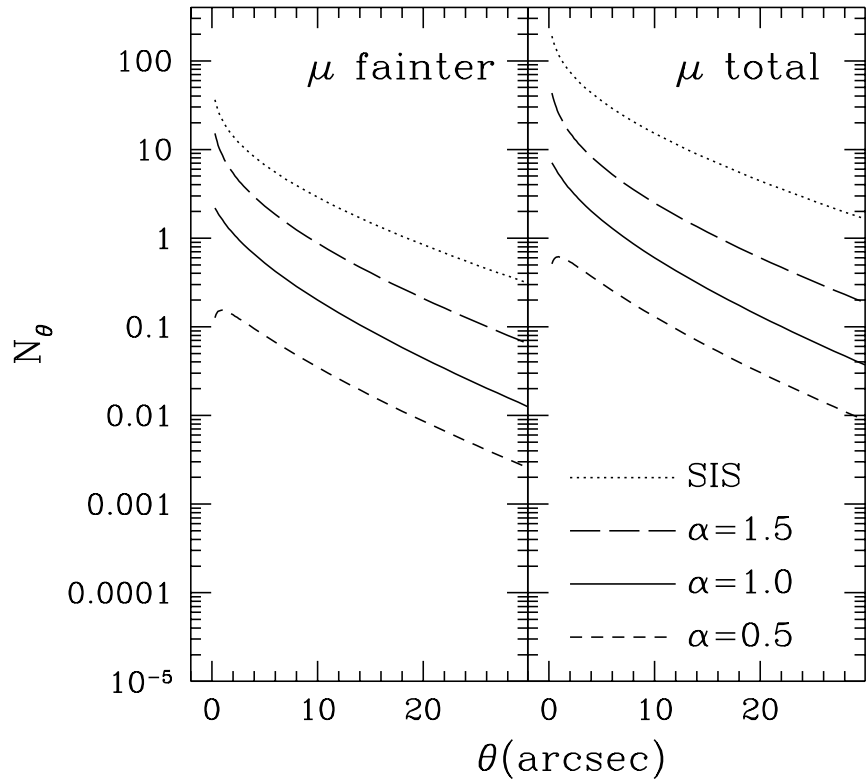}
  \end{center}
\caption{The image separation distribution expected for SDSS data for 
various lens models. $\Lambda$ model is assumed for cosmology.
 The total QSOs number is estimated to be 26,000.}

\end{figure}

\begin{table}
  \begin{center}
  \setlength{\tabcolsep}{3pt}
  \begin{tabular}{|l|c|c|c|r|} \hline
   &SIS &$\alpha=1.5$ &$\alpha=1$ &$\alpha=0.5$  \\  \hline
 $N_{\theta}(6^{\prime \prime} \leq \theta \leq 30^{\prime \prime})$
  $\mu$ total& 197.5 & 31.0 & 7.3 & 1.6    \\  \hline
  $\mu$ fainter& 37.5 & 10.9 & 2.4 & 0.43 \\  \hline
  \end{tabular}
  \end{center}
\caption{The expected number of large image separation lenses
 $(6^{\prime \prime} \leq \theta \leq 30^{\prime \prime})$ for SDSS data
 for $\Lambda$ model with magnification bias of the fainter image
 (bottom) and the total images (top).
The total QSOs number is expected to be about 26,000.}
\end{table}%

\end{document}